\documentclass[conference]{IEEEtran}

\pdfoutput=1
\usepackage{algorithm}
\usepackage{algorithm,algpseudocode}
\usepackage{amsmath}    
\usepackage{amssymb}
\usepackage{graphicx}   
\usepackage{verbatim}   
\usepackage{color}      
\usepackage{subfig}  
\usepackage{url}
\usepackage{mathtools}
\newcommand{\stkout}[1]{\ifmmode\text{\sout{\ensuremath{#1}}}\else\sout{#1}\fi}
\usepackage{soul} 
\usepackage{kantlipsum}
\usepackage{comment}
\allowdisplaybreaks

\def \ConfigSet {\Gamma}
\def \ConfigIdx {\gamma}

\def \deltafic {\delta_{fi}^{\SFCidx}}

\def \ncore {n^{\textsc{core}}}
\def \ncoref {n^{\textsc{core}}_{f}}

\def \NFV {\textsc{nfv}}

\def \ServiceChainSet {C}
\def \SFCidx {c}
\def \SFCset {C}
\def \ServiceChainIdx {c}
\def \xvic {x_{v}^{\SFCidx i}}
\def \xone {x_{v_s}^{\SFCidx, 1}}
\def \xvone {x_{v}^{\SFCidx, 1}}

\def \zRMP {z_{\ConfigIdx}}
\def \myspace {\hspace*{-1.cm} }

\def \SCCopies {Instances}

\def \SCcopies {instances}
\def \SCcopy {instance}

\begin{document}
%
\title{Service Chain (SC) Mapping with Multiple SC {\SCCopies} in a Wide Area Network\\
       \large {\color{green}This is a preprint electronic version of the article 
		submitted to IEEE GlobeCom 2017}}

\author{\IEEEauthorblockN{Abhishek Gupta\IEEEauthorrefmark{1},
Brigitte Jaumard\IEEEauthorrefmark{2}, Massimo Tornatore\IEEEauthorrefmark{1}\IEEEauthorrefmark{3}, and
 Biswanath Mukherjee\IEEEauthorrefmark{1}}
\IEEEauthorblockA{\IEEEauthorrefmark{1}University of California, Davis, USA \ \ 
\IEEEauthorrefmark{2}Concordia University, Canada \ \
\IEEEauthorrefmark{3}Politecnico di Milano, Italy \\
Email: \IEEEauthorrefmark{1}\{abgupta,mtornatore,bmukherjee\}@ucdavis.edu  \IEEEauthorrefmark{2}bjaumard@cse.concordia.ca  \IEEEauthorrefmark{3}massimo.tornatore@polimi.it} }

\maketitle

\begin{abstract}
Network Function Virtualization (NFV) aims to simplify deployment of network services by running Virtual Network Functions (VNFs) on commercial off-the-shelf servers. Service deployment involves placement of VNFs and in-sequence routing of traffic flows through VNFs comprising a Service Chain (SC). The joint VNF placement and traffic routing is usually referred as SC mapping. 
In a Wide Area Network (WAN), a situation may arise where several traffic flows, generated by many distributed node pairs, require the same SC, one single {\SCcopy} (or occurrence) of that SC might not be enough. SC mapping with multiple SC {\SCcopies} for the same SC turns out to be a very complex problem, since the sequential traversal of VNFs has to be maintained while accounting for traffic flows in various directions.  

Our study is the first to deal with SC mapping with multiple SC {\SCcopies} to minimize network resource consumption. Exact mathematical modeling of this problem results in a quadratic formulation.
We propose a two-phase column-generation-based model and solution in order to get results over large network topologies within reasonable computational times. Using such an approach, we observe that an appropriate choice of only a small set of SC {\SCcopies} can lead to solution very close to the  minimum bandwidth consumption.   
\end{abstract}


%

\section{Introduction}
\label{intro}

Today's communication networks deploy network services through proprietary hardware appliances (e.g., network functions such as  firewalls, NAT, etc.) which are statically configured. With rapid evolution of applications, however, networks require more agile and scalable service deployment. 

Network Function Virtualization (NFV) \cite{etsi_nfv} offers a solution for more agile service deployment. NFV envisions hardware functionality as software modules called Virtual Network Functions (VNFs). VNFs can be run on commercial-off-the-shelf hardware such as servers and switches in datacenters (DCs), making service deployment agile and scalable. 

\begin{figure*}
  \centering
  \begin{tabular}{cc}
     \subfloat[][One single SC {\SCcopy} mapping: request $r_3$ and $r_2$ take longer paths]{\label{fig:a}\includegraphics[width=.55\textwidth, scale=1]{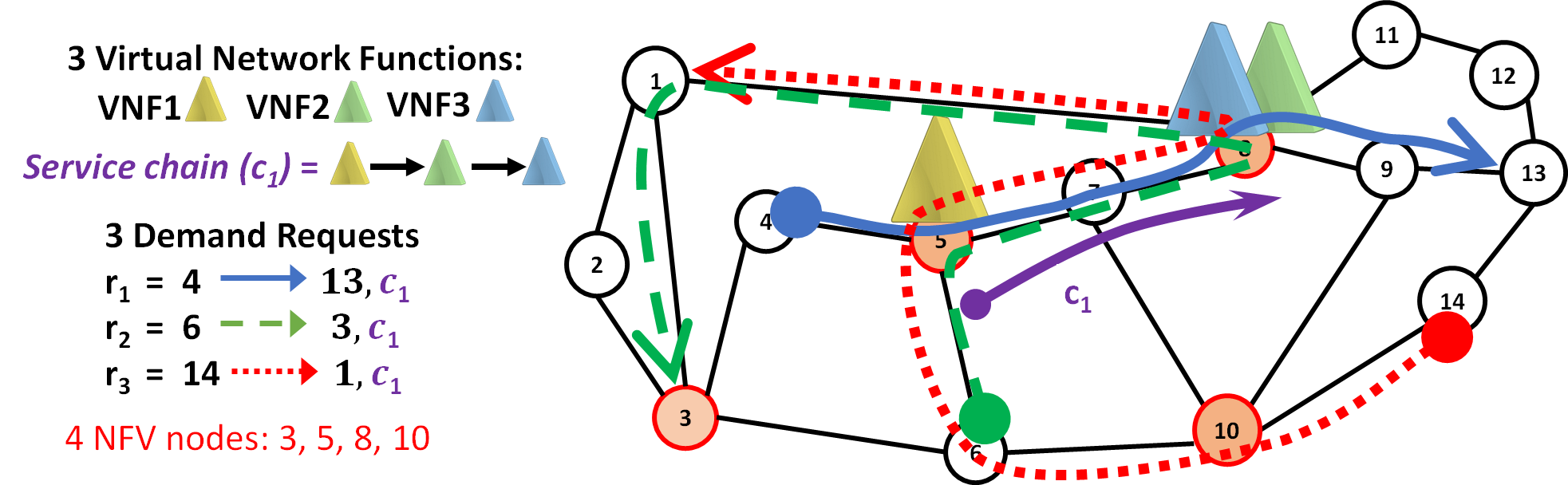}}
    & \subfloat[][Two SC {\SCcopies} mapping: request $r_3$ and $r_2$ take shorter paths]{\label{fig:b}\includegraphics[width=.36\textwidth, scale=1]{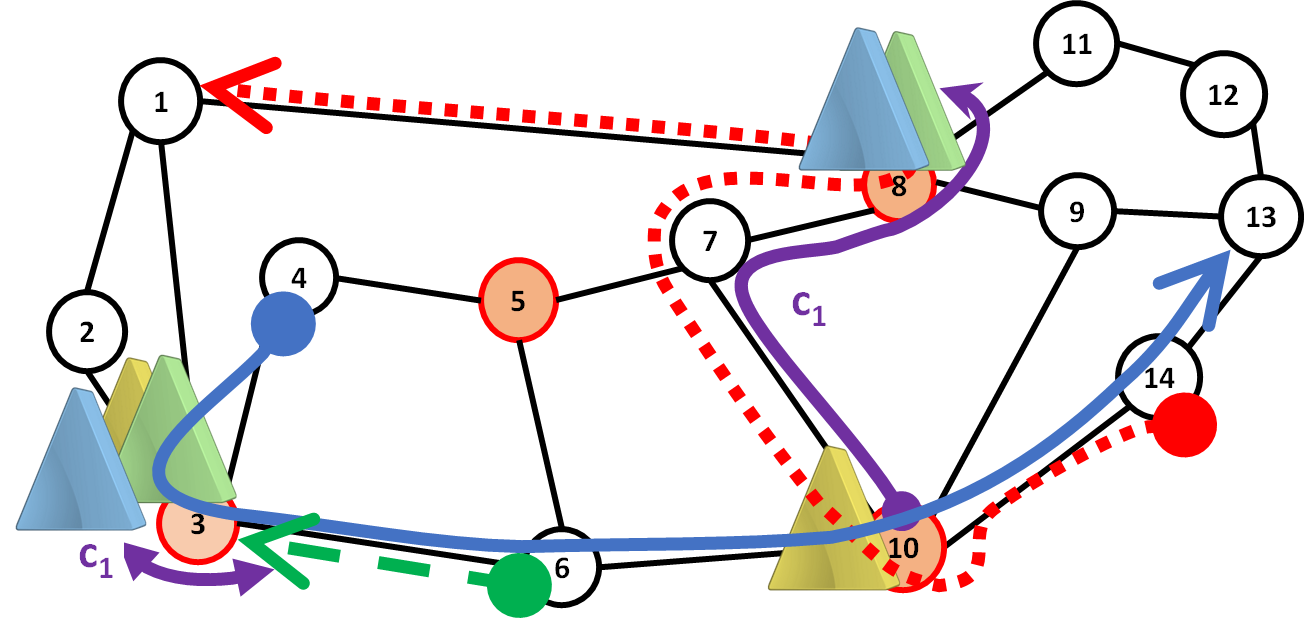}}    
  \end{tabular}
  \caption{Deploying more SC occurrence mappings reduces network resource consumption.}
  \label{configIntro}
\end{figure*}

When several network functions are configured to provide a service, we have a ``Service Chain''. The term ``service chain'' is used ``to describe the deployment of such functions, and the network operator's process of specifying an ordered list of service functions that should be applied to a deterministic set of traffic flows'' \cite{ietf_sc}. So, a ``Service Chain'' (SC) specifies a set of network functions configured in a specific order. With NFV, we can form SCs where VNFs are configured in a specific sequence that allows the minimization of the bandwidth usage in the network (an example is discussed in the next paragaph).

Unfortunately, the fact that VNFs in a single SC need to be traversed by several distinct  traffic flows (i.e., flows requested by multiple geographically-distributed node pairs) in a specific sequence makes it difficult to improve network resource utilization. As an example, let us refer to Figs. \ref{configIntro}\subref{fig:a} and \ref{configIntro}\subref{fig:b}, where three traffic requests $r_1$ (from node 4 to 13), $r_2$ (from node 6 to 3), and, $r_3$ (from node 14 to 1) demand SC $c_1$ composed of VNF1, VNF2 and VNF3 (to be traversed in this order VNF1, VNF2 and VNF3).
In Fig. \ref{configIntro}\subref{fig:a}, we see that if we consider only one mapping occurrence for SC $c_1$, then some traffic flows (in our example, $r_3$ and $r_2$) will be ineffectively routed over a very long  path. Instead, as shown in Fig. \ref{configIntro}\subref{fig:b}, if we use two SC mappings for the same SC, 
we can significantly improve network resource utilization, at the expense of a larger number of VNFs to be deployed/replicated in the network to serve the same SC.
It results in an even more complex problem when in a Wide Area Network (WAN) a very large number of distributed node pairs generate traffic flow, creating a heavily-populated (dense) traffic matrix. Our objective in this work is to reduce the network resource consumption for a WAN with a dense traffic matrix.  


So the question that arises is: how many SC {\SCcopies} for the same SC are required to achieve a quasi-optimal network resource utilization?

A possible trivial solution to the problem of SC mapping in case of multiple node pairs requiring the same SC is to use one single {\SCcopy}, that would most likely lead to host SCs at a single node (e.g., a DC) which is centrally located in the network. 
However, traffic flows will have to take long paths in order to reach the node hosting the SC, which results in a very suboptimal network resource consumption. 

The other extreme case would be to use a distinct SC mapping per each node pair (in other words, the number of SC {\SCcopies} is equal to the number of traffic node pairs). In such a way, we can achieve optimal network resource utilization as each node pair will use an SC  effectively mapped along a shortest path in the network. However,  this approach has the downside of increasing the network orchestration overhead and increase capital expenditure, as there will be a very large number of replicated VNF instances across nodes. To reduce excessive VNF replication, we bound the maximum number of nodes hosting VNFs. Using the shortest path also has the added effect of reducing latency for the service chain, but this aspect is out of scope for this study.  

Intuitively, the number of SC {\SCcopies} for a good solution will be a value between these two extreme points. This solution will minimize the network resource utilization while not excessively increasing the number of nodes hosting VNFs.  

A reasonable trade-off value is difficult to be optimally calculated.
In fact, it has been shown that the problem of SC mapping with multiple SC {\SCcopies} results in quadratic constraints \cite{sc_related}, that severely hamper the scalability of the solution. In this paper, to provide an answer to the question above, we propose a two-phase solution, relying on an ILP column-generation-based model, which provides quasi-optimal solutions with reasonable computational times. Sub-optimality comes from the fact that we solve the problem in two phases: in the first phase we group node pairs that will be forced to use the same SC {\SCcopy}, in the second phase we run our scalable column-generation approach to find a solution starting from the grouping already performed in the first phase.   
Applying the proposed approach over two realistic network topologies, we observe that an appropriate choice of only a small set of different SC mappings can lead to a solution very close to the  minimum theoretical bandwidth consumption, even for a full-mesh traffic matrix. 




The rest of this study is organized as follows. 
Section \ref{relWrk} overviews the existing literature on the SC mapping problem and remarks the novel contributions of this study. Section \ref{probDesc} formally describes the problem and its input parameters.  
Section \ref{sec:sptg} describes a heuristic to cluster groups of node pairs that will use the same SC {\SCcopy}. 
We then describe our  column-generation-based solving method in Section \ref{colGen}. 
Section \ref{sim_examples} provides some illustrative examples that demonstrate  that  a limited number of SC {\SCcopies} can lead to quasi-optimal solution of the problem. Section \ref{concl} concludes the study. 
  

\section{Related Work}
\label{relWrk}

A number of studies exist on the VNF placement and routing problem. 
Mehraghdam \textit{et al.} \cite{sc_related} were the first to formally define the problem of VNF placement and routing. However, they developed a Quadratic Constrained Program (QCP), making it unscalable beyond small problem instances. The authors of Ref. \cite{vnf_placement_turck} study a hybrid deployment scenario with hardware middleboxes using an Integer Linear Program (ILP), but do not enforce VNF service chaining  explicitly. Ref. \cite{place_vnf_secci} uses an ILP to study trade-offs between legacy and NFV-based traffic engineering but does not have explicit VNF service chaining. Ref. \cite{vnf_placement_barcellos_gaspary} models the problem in a DC setting using an ILP to reduce the end-to-end delays and minimize resource over-provisioning while providing a heuristic to do the same. Here too the VNF service chaining is not explicitly enforced by the model. Ref. \cite{orc_vnf_boutaba} models the batch deployment of multiple chains using an ILP and develops heuristics to solve larger instances of the problem. However, they enforce that VNF instances of a function need to be on a single machine and restrict all chains to three VNFs. Our model does not impose such constraints, and we allow any VNF type to be placed on any node and any number of VNFs in a SC while service chaining VNFs for a SC explicitly. Ref. \cite{sc_detail} accounts for the explicit service chaining of VNFs but focuses on compute resource sharing among VNFs. Ref. \cite{nicolas_vnf} also uses a column generation model to solve the VNF placement and routing but considers dedicated SC instances per each traffic pair, hence solving the second extreme case mentioned in the introduction, which is a particular case of our approach.

 
Our previous work \cite{ilp_report} and most of existing works solve the problem for multiple SCs but a single {\SCcopy} of the SC. We remark again that in the current work we consider multiple SCs, but with multiple {\SCcopies} per SC, hence most of the existing works represent a particular case of our current work where, each node pair requesting an SC has its own {\SCcopy}. Further, we also consider multiple geographically distributed node pairs which make a heavily-populated (dense) traffic matrix. Unfortunately, scaling the model to multiple {\SCcopies} per SC results in quadratic constraints. Hence, we now propose a novel decomposition model (column generation) for SC mapping with multiple SC {\SCcopies} (Section \ref{sec:sptg}), that, 
together with a traffic-grouping  heuristic, allows the solution of the problem for multiple same-SC  requests in the same SC {\SCcopy}  (Section \ref{colGen}). 
Our objective is to minimize network-resource consumption  while keeping a bound on the number of nodes that can host VNFs. 

To the best of our knowledge, this is the first attempt to address the solution of the complete SC mapping problem (i.e., with multiple SC {\SCcopies}) over large network instances.


\section{Problem Description}
\label{probDesc}

An operator's network provides multiple services and each service is realized by traversing a Service Chain (SC). Here, we assume that, for each service, the operator knows the ordered sequence of VNFs forming SC. To provide multiple services, a network operator has to map corresponding SCs into network. This is the problem of multiple SC mapping. Most recent works on this topic assume SC mapping with a single SC {\SCcopy} for each SC i.e., all demands for a SC will be mapped to a single {\SCcopy}. Our work takes this further by allowing SC mapping to multiple SC {\SCcopies}, for each SC, which  becomes a more complex problem. We solve this problem by using a two phases approach that will be explained in this the following sections. 

\subsection{Problem Statement}
\label{probState}

Given a network topology, capacity of links, a set of network nodes with NFV support (NFV nodes), compute resources at NFV nodes, the maximum number of NFV-nodes that can be used, traffic flows for source-destination pairs requiring a specific SC with a  certain bandwidth demand, set of VNFs, and, set of SCs, we determine the placement of VNFs and corresponding traffic routing to minimize network-resource (bandwidth) consumption. Note that VNFs can be shared among different SCs.

\subsection{Input Parameters}

\begin{itemize}
\item $G = (V, L)$: Physical topology of backbone network; $V$ is set of nodes and $L$ is set of links.
\item $V^{\NFV} \subseteq V$: Set of NFV nodes.
\item $K$: number of NFV nodes allowed to host VNFs.
\item $F$, indexed by $f$: Set of VNFs.
\item $\ncore$: Number of CPU cores present per NFV node.
\item $\ncoref$: Number of CPU cores per Gbps for function $f$.
\item $\ServiceChainSet$: Set of chains, indexed by $c$.
\item $n_c$: Number of VNFs in SC $\ServiceChainIdx$.
\item $\mathcal{SD}$: Set of source-destination $(v_s,v_d)$ pairs.
\item $D^c_{sd}$: Traffic demand between $v_s$ and $v_d$ for SC $\ServiceChainIdx$.
\item $\sigma_i(\ServiceChainIdx)$: ID of $i$th VNF in SC $\ServiceChainIdx$ where $f_{\sigma_i(\ServiceChainIdx)} \in F$.
\end{itemize}


As already mentioned,  we are solving this problem considering that each SC request can map to multiple {\SCcopies}, which makes the problem quadratic. To avoid quadratic constraints, our solution approach has two phases:
\begin{itemize}
    \item Phase 1: We fix the number $N_c$ of {\SCcopies} accepted per SC ($N_c$ can go from 1 up to the number of demands for that SC), and then we group the traffic requests in $N_c$ groups of requests. All the requests in a group are forced to use the same SC {\SCcopy} (Section \ref{sec:sptg}). Then we pass the $N_c$ {\SCcopies} as distinct SCs to the next phase.
    \item Phase 2: We solve the SC mapping problem with one single {\SCcopy} per SC based on the the inputs of Phase 1. The solution of this simplified (linear, yet still very complex) problem  is based on a column-generation-based decomposition model (Section \ref{colGen}). 
\end{itemize}

\section{Phase 1: Shortest Path Traffic Grouping (SPTG) Heuristic }
\label{sec:sptg} 

In this section, we propose a Shortest Path Traffic Grouping (SPTG) heuristic, which forms $N_c$ groups of node pairs for each SC (given by $SD_c$), to be given as input to  the decomposition model in Section \ref{colGen} that will treat them as distinct SC and  decide the best SC mapping for each of the $N_c$ node-pair groups. As a result, we will have a solution mapping multiple SC {\SCcopies} per SC.  

The logic of the SPTG algorithm below is that groups are formed among node pairs that share links along their shortest path(s). SPTG uses the links on the shortest paths of node pairs to group node pairs together.

\def \clusterG {\textsc{group}}
\def \groupSD {\textsc{cluster}_{sd}}

\begin{algorithm}
\caption{SPTG$(c)$}\label{algo:sptg}
\begin{algorithmic}[1] 
\Require $G$, $SD_c$, $N_c$
\Ensure \textsc{partition} $\leftarrow$ partition of node pairs $(v_s,v_d)$ into groups
\State \textsc{partition} $\leftarrow \emptyset$ 
\State $numberOfGroups \leftarrow 0$
\State $SD_{c}^{\textsc{left}} \leftarrow SD_c$ \Comment{list of $(v_s,v_d)$ for $c$}
\While{$numberOf Groups < N_c \And SD_{c}^{\textsc{left}} \neq \emptyset$}
\For {$(v_s, v_d)$ in $G$}
\State $\groupSD \leftarrow$ set of traffic pairs whose shortest path passes through $(v_s,v_d)$
\EndFor
\State $largestCluster \leftarrow \max\limits_{(v_s, v_d): D_{sd}^c > 0} \groupSD$ 
\State $SD_{c}^{\textsc{left}} \leftarrow SD_{c}^{\textsc{left}} \setminus largestCluster$  \Comment{remove traffic pairs of $largestCluster$ from $SD_{c}^{\textsc{left}}$}
\State Add $largestCluster$ to \textsc{partition} 
\State $numberOfGroups \leftarrow numberOfGroups + 1$
\EndWhile
\If{$SD_{c}^{\textsc{left}} \neq \emptyset$}
\For{$trafficPair \in SD_{c}^{\textsc{left}}$}
\State add $trafficPair$ to  $\clusterG \in$ \textsc{partition}, such that the $(v_s,v_d)$ associated with $\clusterG$ provides the shortest path for provisioning $trafficPair$
\EndFor
\EndIf
\end{algorithmic}
\end{algorithm}



If Algorithm \ref{algo:sptg} terminates with $SD_c^{\textsc{left}}=\emptyset$ and a number of groups that is $< N_c$, partition some of the groups in order to reach $N_c$ groups.

\section{Phase 2: Column-Generation Approach}
\label{colGen}

To present the decomposition a.k.a column generation model, we use the following notation for SC representation. Each SC, denoted by $\ServiceChainIdx$, is characterized by an ordered set of $n_c$ functions:
\begin{equation}
\text{[SC $\ServiceChainIdx$]} \qquad 
f_{\sigma_1(\ServiceChainIdx)} \prec f_{\sigma_2(\ServiceChainIdx)} \prec \dots \prec f_{\sigma_{n_\ServiceChainIdx}(\ServiceChainIdx)}
\end{equation}
Each deployment of SC $\ServiceChainIdx$ is defined by a set of VNF locations, and a set of paths, from the location of first VNF to location of last VNF.

Our decomposition model is based on a set of \textit{SC configurations} where each configuration is associated with a potential provisioning of a SC $\ServiceChainIdx$ and a potential node placement of its functions. 
Let $\ConfigSet$ be the set of configurations, and $\ConfigSet_\ServiceChainIdx$ be the subset of configurations associated with service chain (note that, to keep the problem  linear, only one configuration per SC must be selected)
$\ServiceChainIdx \in \ServiceChainSet$:
$\quad \ConfigSet = \bigcup\limits_{\ServiceChainIdx \in \ServiceChainSet} \ConfigSet_\ServiceChainIdx.$

As the number of potential configurations grows exponentially with network size, we find the problem to fit naturally in the column-generation framework \cite{jaumard_colgen_rwa}. 

Column generation (\textbf{CG}) is a decomposition technique, where the problem (called Master Problem - \textbf{MP}) to be solved is divided into two sub-problems: restricted master problem (\textbf{RMP}) (selection of the best configurations) and pricing problems
 (\textbf{PP\_SC($ \ServiceChainIdx$)})$_{\ServiceChainIdx \in \ServiceChainSet}$ (configuration generators for each chain). 
The CG process involves solving the \textbf{RMP}, querying the dual values of \textbf{RMP} constraints, and using them for \textbf{PP\_SC($ \ServiceChainIdx$)} objective. Each improving solution (i.e., with a negative reduced cost) of \textbf{PP\_SC($ \ServiceChainIdx$)} is added to \textbf{RMP}, and previous step is repeated until optimality condition is reached (see \cite{jaumard_colgen_rwa,jau17RWA}), with the \textbf{PP\_SC($ \ServiceChainIdx$)} explored in a round robin fashion.
A chain configuration is characterized by the following parameters: 
\begin{itemize}
\item Location of the functions: $a_{vi}^{\ConfigIdx} =1$ if $i$th function $f_i \in \ServiceChainIdx$ is located in $v$ in configuration; 0 otherwise.
\item Connectivity of the locations: path from the location of current VNF to next VNF in SC $\ServiceChainIdx$. If link $\ell$ is used in the path from the location of $f_{\sigma_i(\ServiceChainIdx)}$ to the location of $f_{\sigma_{i+1}(\ServiceChainIdx)}$, then $b_{i \ell}^{\ConfigIdx} = 1$; 0 otherwise.
\end{itemize}

\subsection{Restricted Master Problem (\textbf{RMP}) }
\label{masterProb}

\textbf{RMP} selects the best $\ConfigIdx \in \ConfigSet_\ServiceChainIdx$ for each SC $\ServiceChainIdx$. Also it finds a route from $v_s$ (source) to first VNF of $\ServiceChainIdx$ and from last VNF of $\ServiceChainIdx$ to $v_d$ (destination).

\begin{figure}[htb]
\begin{center}
  \subfloat[][A first configuration ($\ConfigIdx_1$) for $c$]{\label{fig:Config1}\includegraphics[scale=.25]{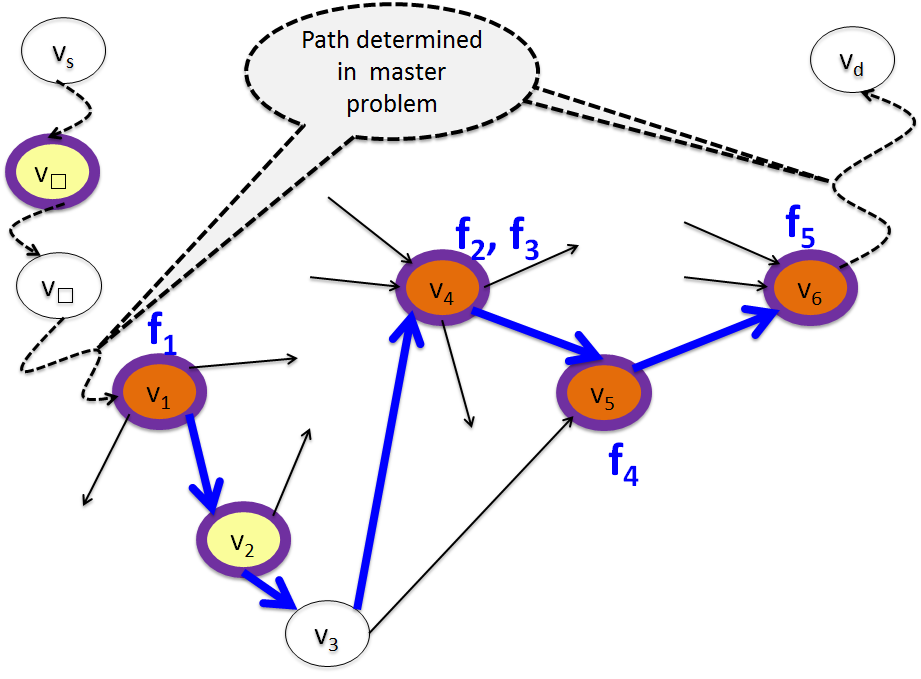}} \\
  \subfloat[][A second configuration ($\ConfigIdx_2$) for $c$]{\label{fig:Config2}\includegraphics[scale=.25]{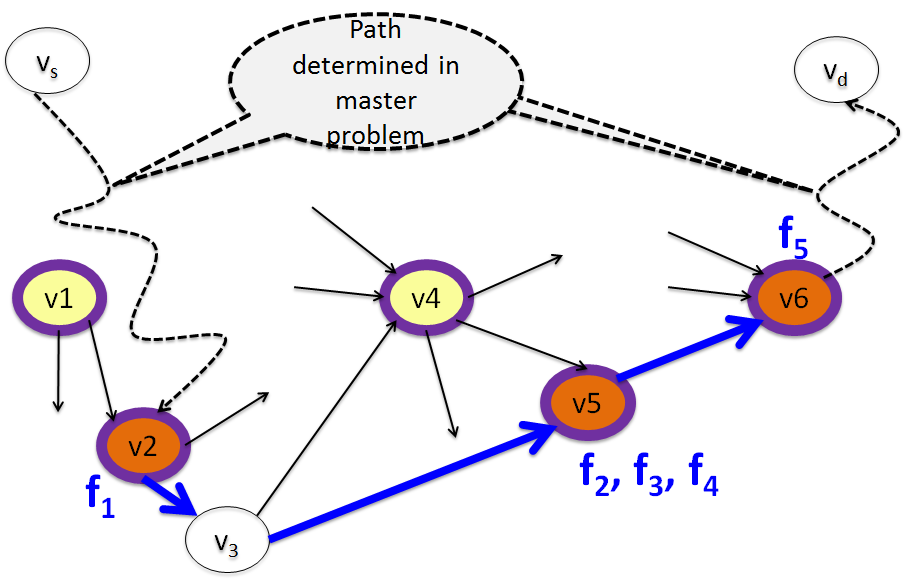}}
\end{center}
\caption{Two configuration examples for  chain $c = (f_1 \prec f_2 \prec f_3 \prec f_4 \prec f_5)$.}
\label{fig:service_chain}	
\end{figure}

An illustration of the constraint splitting between \textbf{RMP} and \textbf{PP\_SC($ \ServiceChainIdx$)} is depicted in Fig. \ref{fig:service_chain}. Nodes circled in purple are NFV nodes, yellow nodes do not host VNFs at present but have NFV support, and orange nodes currently host VNFs. Figure \ref{fig:service_chain}\subref{fig:Config1} has $f_1$ located at $v_1$. When a different configuration is selected in Fig. \ref{fig:service_chain}\subref{fig:Config2} and $f_1$ is located at $v_2$, then \textbf{RMP} finds the path from $v_s$ to location of $f_1$. Similarly, \textbf{RMP} finds the path from last VNF to $v_d$, i.e., $f_5$ to $v_d$ here. 

\vspace{0.2cm}
\noindent
\textbf{Variables:}
\begin{itemize}
\item $z_{\ConfigIdx} =1$ if configuration $\ConfigIdx$ is selected; 0 otherwise.
\item $\xvic =1$ if $i$th function of $\ServiceChainIdx$ is located in $v$; 0 otherwise.
\item $y^{\text{first(\ServiceChainIdx)}, sd}_{\ell} =1$ if $\ell$ is on path from $v_s$ to location of first VNF in $\ServiceChainIdx$; 0 otherwise.
\item $y^{\text{last(\ServiceChainIdx)}, sd}_{\ell} =1$ if $\ell$ is on path from location of last VNF in $\ServiceChainIdx$ to $v_d$; 0 otherwise.
\item $h_v =1$ if $v$ is used as a location for a VNF; 0 otherwise.
\end{itemize}

\vspace{0.2cm}
\noindent
\textbf{Objective:} Minimize bandwidth consumed:
\begin{multline} \min \quad 
     \sum\limits_{\ConfigIdx \in \ConfigSet}  \>
     \underbrace{ 
     \overbrace{ \left(   \sum\limits_{(s,d) \in \mathcal{SD}}   \> D^c_{sd}  \right) }^{\text{Overall traffic using } \ServiceChainIdx} 
     \overbrace{ \left( \sum\limits_{\ell \in L} \> \sum\limits_{i \in I} b_{i \ell}^{\ConfigIdx} \right)}^{ \substack{\text{Number of links} \\  \text{ in the route of } \ServiceChainIdx}}
                   }_{\textsc{cost}_{\ConfigIdx}} 
                   \zRMP  + \\
     \sum\limits_{\ServiceChainIdx \in \ServiceChainSet} \> 
     \sum\limits_{\ell \in L} \> \sum\limits_{(s,d) \in \mathcal{SD}}   \> D^c_{sd}  \left( y^{f_1(\ServiceChainIdx), sd}_{\ell} + y^{f_{n_c}(\ServiceChainIdx), sd}_{\ell} \right).    
\end{multline}

Total bandwidth consumed in placing multiple SCs depends on configuration  $\ConfigIdx$  selected for each SC  $\ServiceChainIdx$. Each $\ConfigIdx$ for $\ServiceChainIdx$ locates VNFs of $\ServiceChainIdx$ and gives the route to traverse these VNF locations. So, bandwidth consumed when going from $v_s$ to $v_d$ and traversing the SC depends on selected $\ConfigIdx$.  


\noindent
\textbf{Constraints}:
\begin{alignat}{2}
 & \sum\limits_{{\ConfigIdx} \in \ConfigSet_\ServiceChainIdx} z_{\ConfigIdx} = 1  
 && \hspace*{-4.cm} \ServiceChainIdx \in \ServiceChainSet \label{eq:single_config_per_service_chain} \\
& \sum\limits_{\SFCidx \in \SFCset} \> \sum\limits_{\ConfigIdx \in  \ConfigSet_{\SFCidx}} 
     \> \sum\limits_{(v_s, v_d) \in \mathcal{SD}}  D_{sd}^c 
     \> (\sum\limits_{i=1}^{n_c}  a_{v i}^{\ConfigIdx} \deltafic  \ncoref) \> \zRMP \leq \ncore 
&& \myspace \nonumber \\
& &&  \hspace*{-4.cm} v \in V^{\NFV} \label{eq:capa_cores} \\
&   \sum\limits_{\ServiceChainIdx \in \ServiceChainSet} \>  \sum\limits_{(v_s, v_d) \in \mathcal{SD}}  \> D^{\ServiceChainIdx}_{sd}  && \nonumber \\
&\qquad  \left(    y^{f_1(\ServiceChainIdx), sd}_{\ell} 
        + y^{f_{n_c}(\ServiceChainIdx), sd}_{\ell}     
        + \sum\limits_{{\ConfigIdx} \in \ConfigSet_{\ServiceChainIdx}} \> \sum\limits_{i =1}^{n_{\ServiceChainIdx} - 1}  b_{i \ell}^{\ConfigIdx} \>    
   \zRMP \right)   && \nonumber \\
& \qquad \qquad \qquad \qquad 
  \leq \textsc{cap}_{\ell}  
&& \hspace*{-2.cm}  \ell \in L \label{eq:capacity} \\
& \sum\limits_{\ConfigIdx \in \ConfigSet_{\ServiceChainIdx}}  a_{v i}^{\ConfigIdx}  \zRMP =  \xvic  
&&   \hspace*{-4.cm} f_i \in F(c), \ServiceChainIdx \in \ServiceChainSet, v \in V^{\NFV} \label{eq:a_x_consistent1} \\
&  M x_{vf} \geq \sum\limits_{\ServiceChainIdx \in \ServiceChainSet: f \in \ServiceChainIdx} \> \sum\limits_{i \in \{1,2,\dots, n_c\}: f_i = f}  \xvic    \geq x_{vf} 
 && \nonumber \\
& && \hspace*{-3.cm} v \in V^{\NFV}, f_i \in F \label{eq:f_on_v} \\
& M h_v \geq \sum\limits_{f \in F} x_{vf} \geq h_v && \hspace*{-3.cm} v \in V^{\NFV} \label{eq:vnf_location} \\
& \sum\limits_{v \in V^{\NFV}} h_v \leq K \label{eq:k_nfv_nodes}.
\end{alignat}  

Constraints \eqref{eq:single_config_per_service_chain} guarantee that we select exactly one $\ConfigIdx$ for SC  $\ServiceChainIdx$  
and forces $\ServiceChainIdx$ to have a single instance. 
Each $\ConfigIdx$ is associated with a set of $a^{\ConfigIdx}_{vi}$ (from \textbf{PP\_SC($ \ServiceChainIdx$)}) required to be consistent with $\xvic$ in \textbf{RMP}, which is resolved by Eqs. \eqref{eq:a_x_consistent1}.

Constraints \eqref{eq:capa_cores} ensure that each NFV node has a sufficient number of CPU cores for hosting $f$. Those constraints also accounts for increase in compute resource due to traffic increase. 
Eq. \eqref{eq:capacity} enforces link-capacity constraints for the complete route for SC  $\ServiceChainIdx$  from $v_s$ to $v_d$  for all $(v_s,v_d) \in \mathcal{SD} : D^{c}_{sd} > 0$).

Eq. \eqref{eq:f_on_v} is used to keep track of VNF replicas. Eq. \eqref{eq:vnf_location} is used to keep track of NFV nodes used for hosting VNFs while Eq. \eqref{eq:k_nfv_nodes} enforces the number of NFV nodes allowed to host VNFs.

%
%

\begin{alignat}{2}     
& \text{\textbf{Route from} } v_s \text{ \textbf{to first function location:}} \nonumber \\
& \sum\limits_{\ell \in \omega^+{(v_s)}} y^{f_1(\ServiceChainIdx), sd}_{\ell} = 1 - \xone
&& \myspace  \ServiceChainIdx \in \ServiceChainSet, \nonumber \\
& &&  \hspace*{-3.5cm}  (v_s,v_d) \in \mathcal{SD}: D_{sd}^{\ServiceChainIdx} > 0 \label{eq:link_from_source_to_ingress} \\
& \sum\limits_{\ell \in \omega^-{(v)}} y^{f_1(\ServiceChainIdx), sd}_{\ell}  \geq \xvone
&&  \hspace*{-2.cm}  \ServiceChainIdx \in \ServiceChainSet, \nonumber \\
& && \hspace*{-5.cm}  (v_s,v_d) \in \mathcal{SD}: D_{sd}^{\ServiceChainIdx} > 0, v \in V^{\NFV} \setminus \{ v_s \} \label{eq:to_ensure_NFV_1_placement} \\
&     \sum\limits_{\ell \in \omega^+{(v)}} y^{f_1(\ServiceChainIdx), sd}_{\ell}
     -  \sum\limits_{\ell \in \omega^-{(v)}} y^{f_1(\ServiceChainIdx), sd}_{\ell} 
     = - \xvone
&& \nonumber \\
& && \hspace*{-6.cm}  \ServiceChainIdx \in \ServiceChainSet, 
            (v_s,v_d) \in \mathcal{SD}: D_{sd}^{\ServiceChainIdx} > 0, v \in V^{\NFV} \setminus \{ v_s \}        \label{eq:places_NFV} \\
&     \sum\limits_{\ell \in \omega^+{(v)}} y^{f_1(\ServiceChainIdx), sd}_{\ell}
     -  \sum\limits_{\ell \in \omega^-{(v)}} y^{f_1(\ServiceChainIdx), sd}_{\ell}
     = 0
      && \nonumber \\
      & && \hspace*{-7.cm}  \ServiceChainIdx \in \ServiceChainSet, (v_s,v_d) \in \mathcal{SD}: D_{sd}^{\ServiceChainIdx} > 0, v \in V \setminus (V^{\NFV} \cup \{ v_s \}).         \label{eq:places_non_NFV} 
\end{alignat} 

We assume that a unique route exists from $v_s$ to first VNF location. This is imposed by selecting exactly one outgoing link from $v_s$ unless first VNF is located at 
$v_s$. We account for these scenarios using Eq. \eqref{eq:link_from_source_to_ingress}. To find the route from $v_s$ to first VNF, flow conservation needs to be enforced at the intermediate nodes which may or may not have NFV support. Eqs. \eqref{eq:places_NFV} and \eqref{eq:places_non_NFV} enforces flow-conservation constraints at nodes with and without NFV support, respectively.

We can enforce same functionality as Eqs. \eqref{eq:link_from_source_to_ingress},  \eqref{eq:places_NFV}, \eqref{eq:places_non_NFV}, and \eqref{eq:to_ensure_NFV_1_placement},  on route from location of last VNF to $v_d$. For the interested reader, similar details are provided in \cite{gupta_icc_arxiv}.

\begin{figure*}
  \centering
  \begin{tabular}{cc}
     \subfloat[][NSFNET K=14]{\label{fig:aa}\includegraphics[width=.5\textwidth, scale=1]{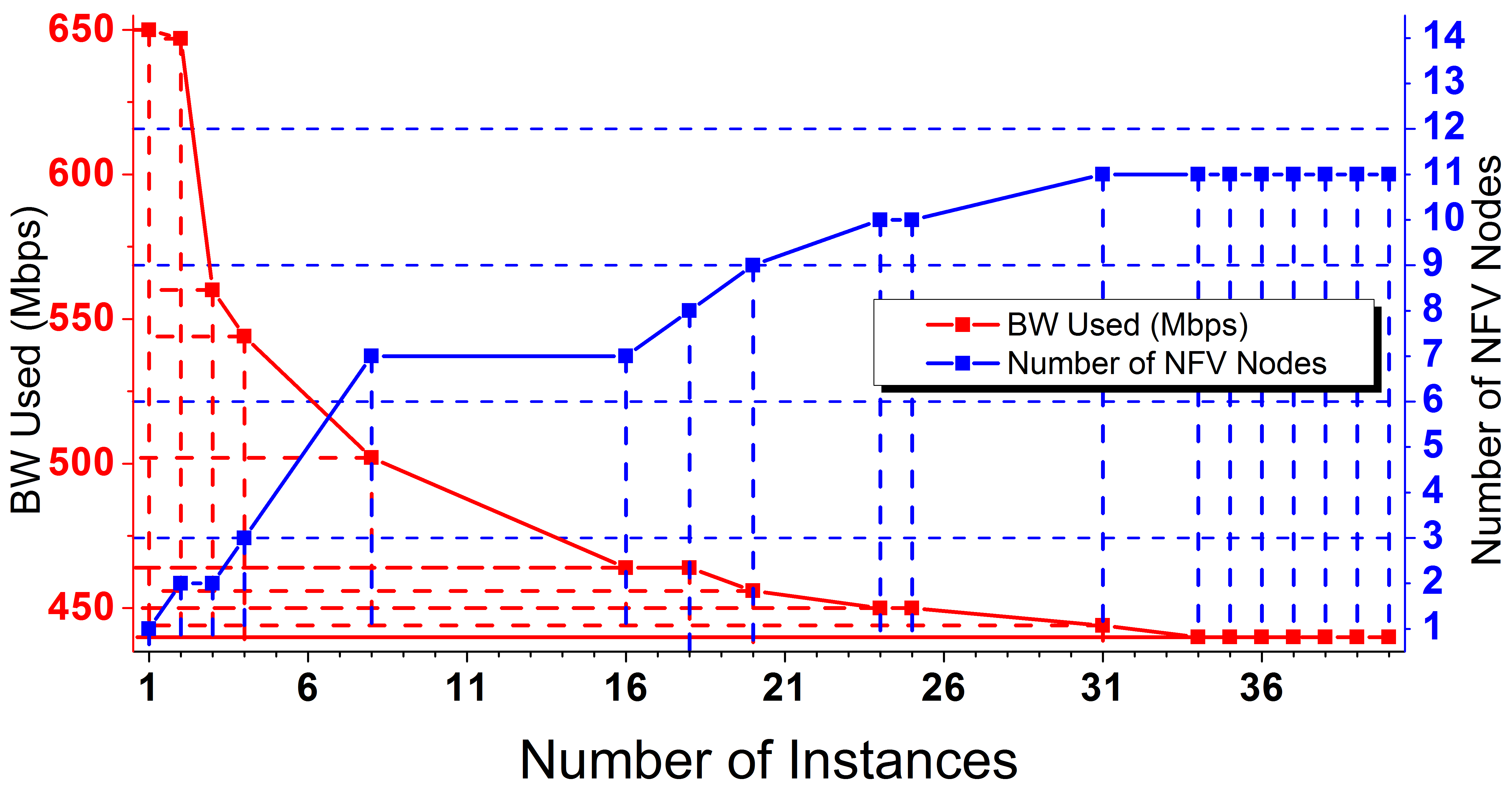}}
    & \subfloat[][NSFNET K=1,2,3,4,5,14]{\label{fig:bb}\includegraphics[width=.45\textwidth, scale=1]{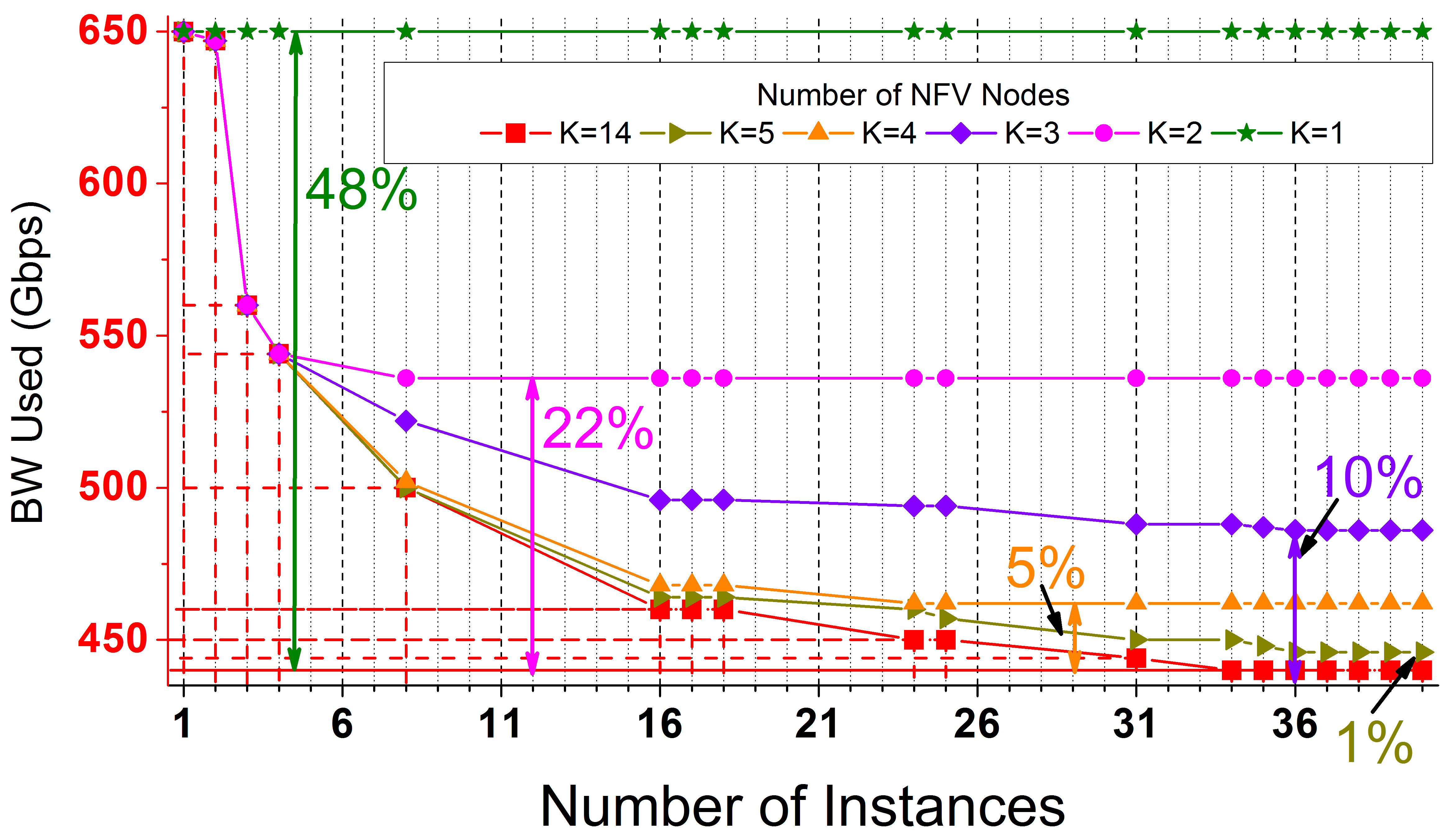}}
    \\ \\
     \subfloat[][COST239 K=11]{\label{fig:cc}\includegraphics[width=.5\textwidth, scale=1]{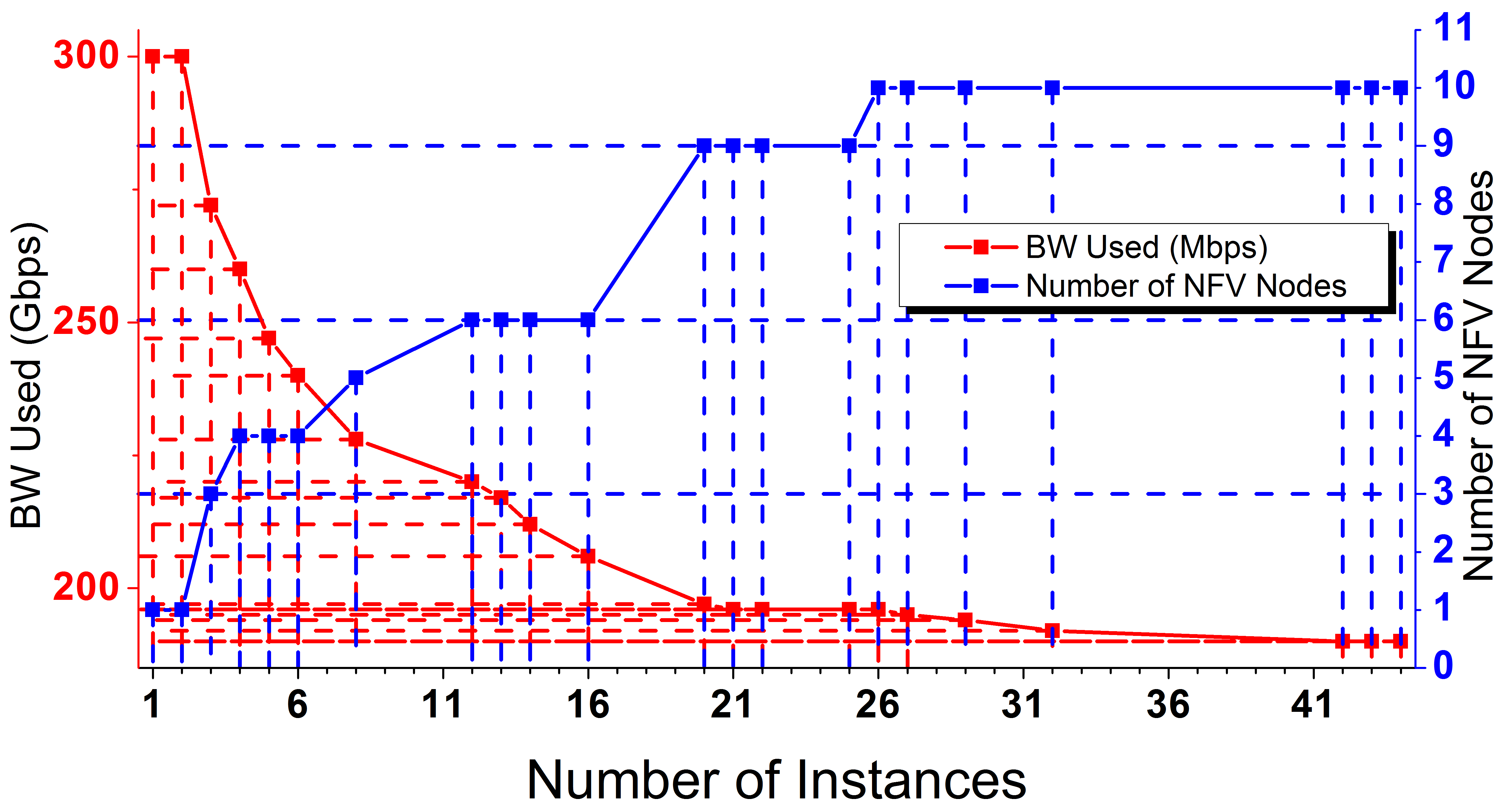}}
    & \subfloat[][COST239 K=1,2,3,4,5,11]{\label{fig:d}\includegraphics[width=.45\textwidth, scale=1]{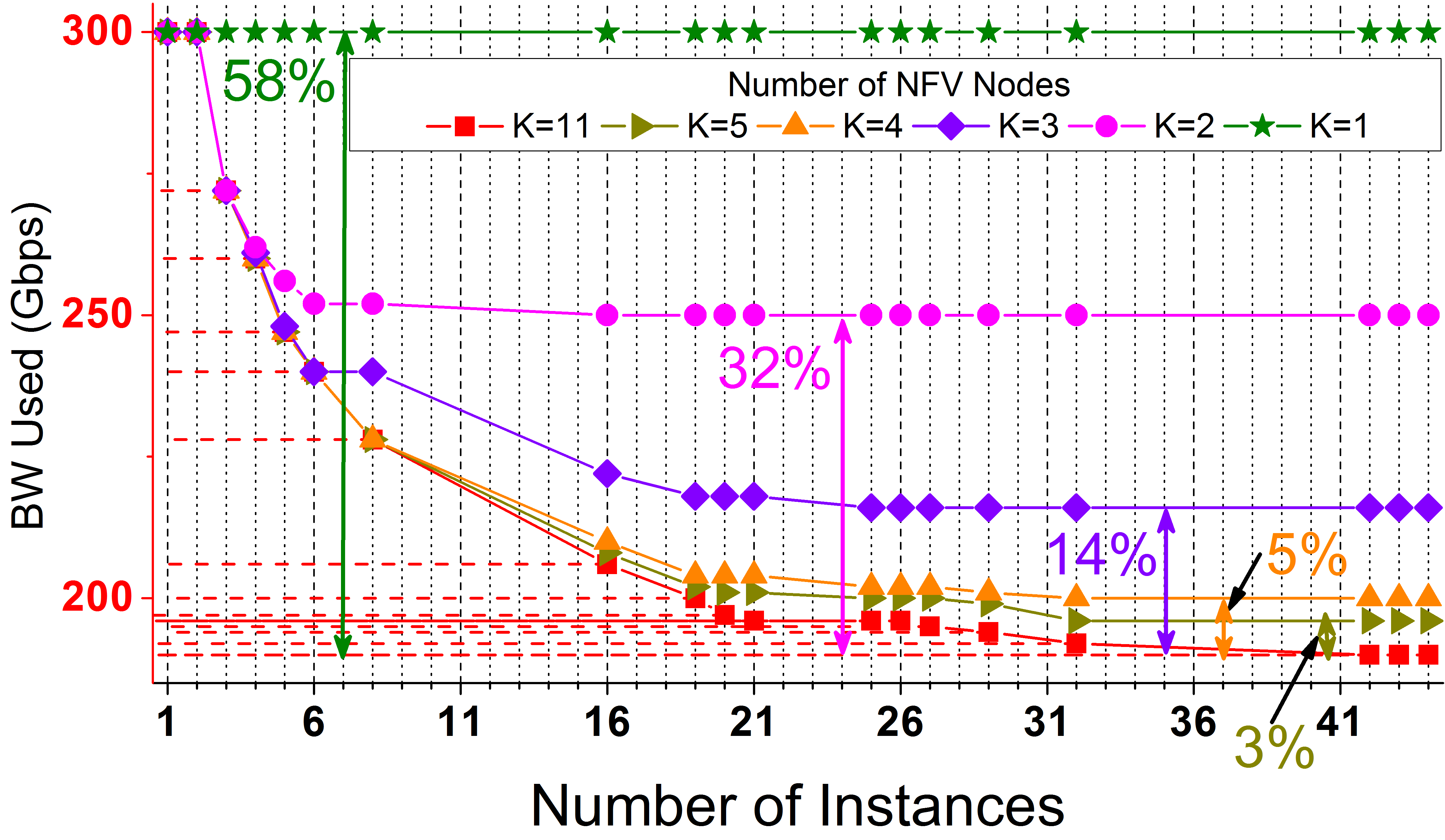}}  
  \end{tabular}
  \caption{Bandwidth vs. Number of NFV Nodes in NSFNET and COST239 networks.}
  \label{results}
\end{figure*}

\subsection{Pricing Problem}
\label{pricingProb}

Mapping computations for each SC  $\ServiceChainIdx$  ($\ServiceChainIdx \in \ServiceChainSet$) correspond to the solution of pricing problems. The number of pricing problems to be solved equals the sum of the number of SC {\SCcopies} to be deployed. 

Pricing problem \textbf{PP\_SC($ \ServiceChainIdx$)} generates: 
\textit{(i)} A set of locations for VNFs of  $\ServiceChainIdx$; and
\textit{(ii)} a sequence of paths from the location of VNF $f_i$ to the location of VNF $f_{i+1}$, for $i = 1, 2, \dots, n_\ServiceChainIdx - 1$ for chain $\ServiceChainIdx$. Each solution that is generated by \textbf{PP\_SC($\ServiceChainIdx$)} with a negative reduced cost, leads to a new potential $\ConfigIdx$ for  $\ServiceChainIdx$ of interest. Please refer to \cite{gupta_icc_arxiv} for similar details.

\subsection{Solution Scheme}

The \textbf{PP\_SC($\ServiceChainIdx$)} are solved in a round-robin fashion and the final \textbf{RMP} is solved as an ILP, as in \cite{jaumard_colgen_rwa,jau17RWA}. 


\section{Illustrative Numerical Examples}
\label{sim_examples}

We first tested our two-phase optimization process on a 14 node NSFNet WAN topology \cite{gupta_icc_arxiv} with a complete traffic matrix, i.e., with traffic flows between all node pairs, assuming all nodes can be made NFV nodes. The link capacity is sufficient to support all flows. Each traffic flow is 1 Gbps and demands the same 3 VNF service chain (SC). Compute resource (CPU) at each node is sufficient to support all VNF placements. The second run of the model is on a 11 node COST239 WAN topology \cite{farhan_cost239} under the same specifications as above.    

Figure \ref{results}\subref{fig:a} shows the bandwidth consumption as the number of SC {\SCcopies} increases. 
Here, we allow all nodes (K=14) to host VNFs. 
We find that as the number of deployed SC {\SCcopies} increases, the bandwidth consumption decreases. Indeed, with a higher number of {\SCcopies}, groups of traffic node pairs are able to take short paths. We see that at 34 {\SCcopies}, we achieve minimum possible bandwidth consumption, meaning traffic flow is taking the shortest paths. 
Note that the number of traffic node pairs in the network is 182, requiring a priori 182 different {\SCcopies} (solving the problem for 182 {\SCcopies} would be equivalent to obtaining a solution with existing models as in Ref. \cite{place_vnf_secci}\cite{vnf_placement_barcellos_gaspary}\cite{orc_vnf_boutaba}). Instead, our approach with 34 {\SCcopies} already achieves optimized bandwidth consumption. This is very important as a lower number of {\SCcopies} lowers the orchestration overhead for network operators. 

Further, the number of NFV nodes increases as the number of SC {\SCcopies} increases. Indeed, as SC mappings become more distributed,  more nodes are being used for hosting virtual functions. In Fig. \ref{results}\subref{fig:aa}  11 nodes are NFV enabled for 34 different SC mappings. 
For a network operator, capital expenditure in making 11 out of 14 nodes capable of hosting VNFs is very high. 
So, operators may want to minimize the number of NFV nodes while also trying to reduce bandwidth consumption by using multiple SC mapping {\SCcopies}. This led us to explore how the bandwidth consumption varies when the numbers of NFV nodes are limited.  

Figure \ref{results}\subref{fig:b} shows the bandwidth consumption for SC mapping {\SCcopies} for various $K$ values. When $K=1$, all traffic flows have to traverse the one node in the network and hence, the number of {\SCcopies} does not affect bandwidth consumption. At $K=2$, deploying more than 8 {\SCcopies} does not improve bandwidth utilization. For $K=3$ and 16 {\SCcopies} we are able to achieve close to 10\% of the minimum bandwidth utilization. Similarly, at $K=4$ and 24 {\SCcopies}, we reach within 5\% of the optimal bandwidth consumption. The bandwidth consumption comes to within 1\% of the optimal when K=5 and 36 {\SCcopies}. Thus, we establish that we can achieve near-to-optimal bandwidth consumption by a using a relatively small number of {\SCcopies} and nodes.

Figures \ref{results}\subref{fig:cc} and \ref{results}\subref{fig:d} corroborate our findings in Figs. \ref{results}\subref{fig:aa} and \ref{results}\subref{fig:bb}. With the COST239 network too, we can achieve near-to-optimal bandwidth consumption with a  small number of SC {\SCcopies} and nodes.

\vspace*{-.5cm}
\section{Conclusion}
\label{concl}

We introduce the problem of multiple service chain (SC) mapping with multiple SC {\SCcopies} in high traffic. 
We developed a column generation model along with a Shortest Path Traffic Grouping (SPTG) heuristic which results in a scalable linear model, thereby solving this complex problem in a relatively small amount of time. Further, we demonstrate that a near-to-optimal network resource consumption can be achieved with a relatively small number of SC {\SCcopies} and NFV nodes. This is critical in order to reduce the network operator's orchestration overhead and capital expenditures. 

\vspace*{-.2cm}
\section*{Acknowledgment}
This work was supported by NSF Grant No. CNS-1217978.



%
\bibliographystyle{IEEEtran}
\bibliography{gupta_nfv}

\begin{thebibliography}{10}
\providecommand{\url}[1]{#1}
\csname url@samestyle\endcsname
\providecommand{\newblock}{\relax}
\providecommand{\bibinfo}[2]{#2}
\providecommand{\BIBentrySTDinterwordspacing}{\spaceskip=0pt\relax}
\providecommand{\BIBentryALTinterwordstretchfactor}{4}
\providecommand{\BIBentryALTinterwordspacing}{\spaceskip=\fontdimen2\font plus
\BIBentryALTinterwordstretchfactor\fontdimen3\font minus
  \fontdimen4\font\relax}
\providecommand{\BIBforeignlanguage}[2]{{%
\expandafter\ifx\csname l@#1\endcsname\relax
\typeout{** WARNING: IEEEtran.bst: No hyphenation pattern has been}%
\typeout{** loaded for the language `#1'. Using the pattern for}%
\typeout{** the default language instead.}%
\else
\language=\csname l@#1\endcsname
\fi
#2}}
\providecommand{\BIBdecl}{\relax}
\BIBdecl

\bibitem{etsi_nfv}
ETSI, ``Network functions virtualisation: Introductory white paper,''
  \url{portal.etsi.org/NFV/NFV_White_Paper.pdf}, 2012.

\bibitem{ietf_sc}
IETF, ``Network service chaining problem statement,''
  \url{https://tools.ietf.org/html/draft-quinn-nsc-problem-statement-00}, 2013.

\bibitem{sc_related}
S.~Mehraghdam, M.~Keller, and H.~Karl, ``Specifying and placing chains of
  virtual network functions,'' in \emph{IEEE International Conference on Cloud
  Networking (CloudNet)}, 2014, pp. 7--13.

\bibitem{vnf_placement_turck}
H.~Moens and F.~De~Turck, ``{VNF-P: A model for efficient placement of
  virtualized network functions},'' in \emph{10th International Conference on
  Network and Service Management (CNSM)}, Nov. 2014, pp. 418--423.

\bibitem{place_vnf_secci}
B.~Addis, D.~Belabed, M.~Bouet, and S.~Secci, ``Virtual network functions
  placement and routing optimization,''
  \emph{https://hal.inria.fr/hal-01170042/}, 2015.

\bibitem{vnf_placement_barcellos_gaspary}
M.~C. Luizelli, L.~R. Bays, L.~S. Buriol, M.~P. Barcellos, and L.~P. Gaspary,
  ``{Piecing together the NFV provisioning puzzle: Efficient placement and
  chaining of virtual network functions},'' in \emph{IFIP/IEEE IM}, May 2015,
  pp. 98--106.

\bibitem{orc_vnf_boutaba}
\BIBentryALTinterwordspacing
M.~Bari, S.~Chowdhury, R.~Ahmed, and R.~Boutaba, ``On orchestrating virtual
  network functions in {NFV},'' \emph{Computing Research Repository}, vol.
  abs/1503.06377, 2015. [Online]. Available:
  \url{http://arxiv.org/abs/1503.06377}
\BIBentrySTDinterwordspacing

\bibitem{sc_detail}
M.~Savi, M.~Tornatore, and G.~Verticale, ``Impact of processing costs on
  service chain placement in network functions virtualization,'' in \emph{IEEE
  NFV-SDN}, 2015.

\bibitem{nicolas_vnf}
\BIBentryALTinterwordspacing
N.~Huin, B.~Jaumard, and F.~Giroire, ``{Optimization of Network Service Chain
  Provisioning}.'' [Online]. Available: \url{https://hal.inria.fr/hal-01476018}
\BIBentrySTDinterwordspacing

\bibitem{ilp_report}
A.~Gupta, M.~Habib, P.~Chowdhury, M.~Tornatore, and B.~Mukherjee, ``{Joint
  Virtual Network Function Placement and Routing of Traffic in Operator
  Networks},'' \emph{Technical Report, UC Davis}, 2015.

\bibitem{jaumard_colgen_rwa}
B.~Jaumard, C.~Meyer, and B.~Thiongane, ``On column generation formulations for
  the {RWA} problem,'' \emph{Discrete Applied Mathematics}, vol. 157, pp.
  1291--1308, 2009.

\bibitem{jau17RWA}
B.~Jaumard and M.~Daryalal, ``Efficient spectrum utilization in large scale
  {RWA} problems,'' \emph{IEEE/ACM Transactions on Networking}, pp. 1 -- 16,
  2017.

\bibitem{gupta_icc_arxiv}
A.~Gupta, B.~Jaumard, M.~Tornatore, and B.~Mukherjee, ``{Multiple Service Chain
  Placement and Routing in a Network-enabled Cloud},'' \emph{CoRR}, vol.
  abs/1611.03197, 2016.

\bibitem{farhan_cost239}
M.~F. Habib, M.~Tornatore, M.~De~Leenheer, F.~Dikbiyik, and B.~Mukherjee,
  ``{Design of disaster-resilient optical datacenter networks},'' \emph{Journal
  of Lightwave Technology}, vol.~30, no.~16, pp. 2563--2573, 2012.

\end{thebibliography}

\end{document}